\documentclass[12pt]{article}
\usepackage{epsfig,graphics,color}
\author{\large \bf  Z.~Usubov\footnote
         {On leave of absence from Institute of Physics, Baku, Azerbaijan}
\\
\\Dzhelepov Laboratory of Nuclear Problems,
\\Joint Institute for Nuclear Research,
\\ 141980, 6 Joliot-Curie,  Dubna, Moscow Region,                  
Russia }

\title { LHC Probe of Quark Substructure at the Early
Stages of Running}

\begin{document}
\maketitle
{
\vskip 0.5cm
\hskip 6.5cm {\bf \large { Abstract}}
\vskip 1.0cm

We explore the opportunity to look for quark
compositeness in the early stages of the LHC running
by analyzing high-$E_T$ dijet production.
The quark substructure  that will  manifest 
itself     by affecting various kinematic                    
distributions at the center-of-mass
energy  $E_{cm}$=10 TeV and in     the integrated
luminosity range from 0.5 to 10 fb$^{-1}$ is investigated.
The estimation of the 
characteristic energy scale $\Lambda$, based on the novel, potentially
powerful observable, 
is  found to be sensitive    
to the sign of the interference between Standard Model and 
four-fermion contact interactions.
We find  the lower limits on  
$\Lambda$ at the LHC to be
about 10.5(15.0) TeV for constructive interference and
8.5(11.5) TeV for destructive one at the integrated luminosity
0.5(10) fb$^{-1}$ and $E_{CM}=10\,$TeV.

\section{Introduction}
$ $

Tremendous effort in the LHC experiments will be
made      to reveal   the origin of 
electroweak symmetry breaking.                
The agent  of the symmetry breaking,
as is formulated in the Standard Model (SM), is a scalar field
giving rise   to one physical neutral scalar Higgs boson. 
On the other hand, the electroweak symmetry breaking can also
arise from QCD-like interacting sector of composite particles
(with a fermion antifermion condensate $< \bar \psi \psi > \neq 0 $
breaking the symmetry).
Indeed, the existence of too many free parameters in the SM
and  three families of quarks and leptons 
could be an indication of the substructure of these objects.

At the energies much below the characteristic energy scale                  
$\Lambda$ with the neglect of the underlying strong dynamics
a new coupling among quarks can be 
approximated by a four-fermion contact interactions.
A search for   evidence  for the quark substructure is usually  based
on the concept of their higher dimensional operator
contribution leading to the contact   interaction
term  in the effective Lagrangian of the form\cite{hinc}
\begin{equation}
    L_{qq}= { \lambda { g^2 \over {2 \, {\Lambda}^2}}
{[\bar q   {\gamma}^{\mu} q] [\bar q   {\gamma}_{\mu} q] }}    
\end{equation}
 where $\lambda =-1(+1)$ defines constructive (destructive)
interference between  the contact and SM
interactions and $g^2$ is the quark compositeness coupling constant.
The compositeness energy scale $\Lambda$ can be chosen such
that ${g^2 \over 4\,\pi}= 1$, and the model is completely
determined by specifying the parameters $\lambda$ and $\Lambda$.

A typical signal of quark compositeness would be an excess of the 
high-$E_T$ jets over the level predicted by QCD and/or
a more isotropic dijet angular distribution than what is
expected from the SM.

The current experimental lower limits on the quark compositeness 
scale from lepton-lepton, lepton-nucleon and nucleon-nucleon
interactions vary from 2.5 to 6.3 TeV\cite{abbi,adlo,gree}, 
depending on the chirality channels under
consideration.                                                        

The expected sensitivity of the ATLAS  experiment to the discovery
of the contact interactions can be found in\cite{tdr}.
Prospects for a luminosity and energy upgraded LHC are also
presented\cite{gian1}.
The possibility of the composite top quark manifestation
at the LHC is now widely discussed\cite{ctop}.

The rest of this note is organized as follows.
In Section~2 we describe all  the observables that
we use to study the quark                 
compositeness. Section~3 gives a brief description
of the data simulation technique. Section~4 gives a
summary of our analysis itself. We  examine the 
effect of the quark substructure on 
the measurables  and the sensitivity of the data 
to the quark compositeness scale.
The sensitivity of the results to the parton
distribution functions (pdfs) is 
also demonstrated in Section~4.
Section~5 presents our conclusions.

\section{Quark substructure observables at the LHC}
$ $

The LHC allows one to reach very large values of jet                   
transverse energy ($E_T>$1.5 TeV) and dijet invariant mass
($m_{j1j2}>$3.0 TeV) even in the early stages of operation.
Such a kinematic region of dijet production has never been
studied before.

\begin{figure}[ht]
\vskip -3.3cm
\centerline{\epsfxsize 6.0 truein \epsfbox{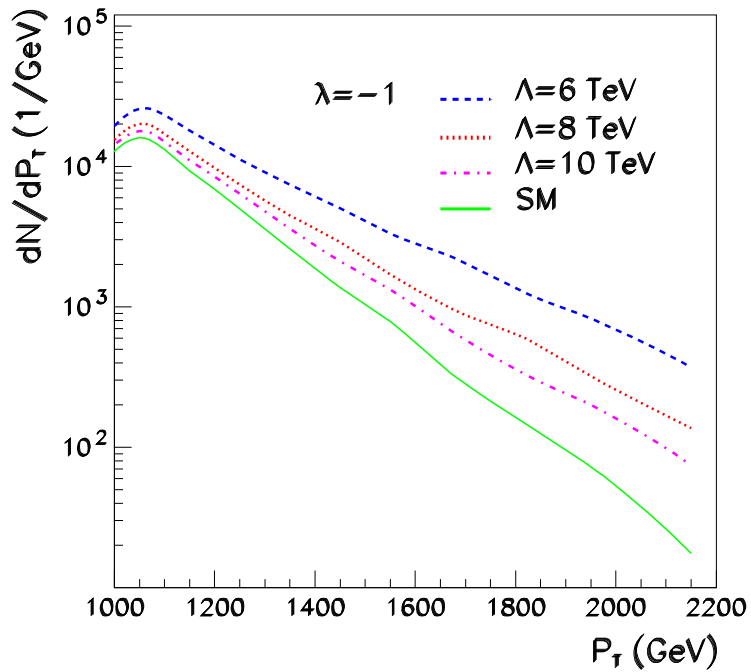}}
\vskip -2.0cm
\caption{$P_T$ distribution of two leading jets for the Standard 
Model and the composite    
quark predictions at different energy scales $\Lambda$ and $E_{CM}=10\,$TeV.
The data correspond to the 
constructive interference between the Standard Model and the contact interactions.}
\end{figure}

To study  dijet properties  in $pp$ collisions at the center-of-mass
energy $E_{CM}=10\,$TeV,  we chose four   variables:     

$\bullet$ $P_T$ distribution of  two most energetic jets;

$\bullet$ $\chi$, defined as $ \chi \equiv \exp{|{\eta}_1-{\eta}_2|}$,
where ${\eta}_{1,2}$ are the pseudorapidities of  two
leading jets. For the case of $2 \to 2$ parton scattering 
$\chi$ is related to the center-of-mass scattering angle ${\theta}^*$
as   
\begin{equation}
   \chi = { {1+|cos{\theta}^*|} \over {1-|cos{\theta}^*|}};   
\end{equation}

$\bullet$ $\alpha_{R}$, which was   proposed by L.~Randall and
D.~Tucker-Smith\cite{rand} as the ratio of     
the second hardest jet transverse momentum and the 
invariant mass  $m_{j1j2}$ of two hardest jets
\begin{equation}
   \alpha_{R} \equiv {{P_{T_2}} \over {m_{j1j2}}};       
\end{equation}

$\bullet$ the dimensionless variable  $\alpha_{Z}$, which                            
we define as $E_{CM} \over 2$ times the 
ratio of the sum  and the product of the                                           
transverse momenta of two hardest jets.
\begin{equation}
   \alpha_{Z} \equiv { E_{CM}  \over 2} \,\,{  { P_{T_1}+P_{T_2} } 
            \over { P_{T_1}\cdot P_{T_2}}}.
\end{equation}  

The variable $\chi$ is defined through angular quantities, and hence         
is less susceptible to the precise knowledge of jet energy
scale but it is sensitive
to the gluon radiation and higher-order corrections.
The use   of the $\chi$ as a dijet angular distribution
variable makes the comparison with theory 
more straightforward\cite{barg}.

\begin{figure}[ht]
\vskip -1.0cm
\centerline{\epsfxsize 5.0 truein \epsfbox{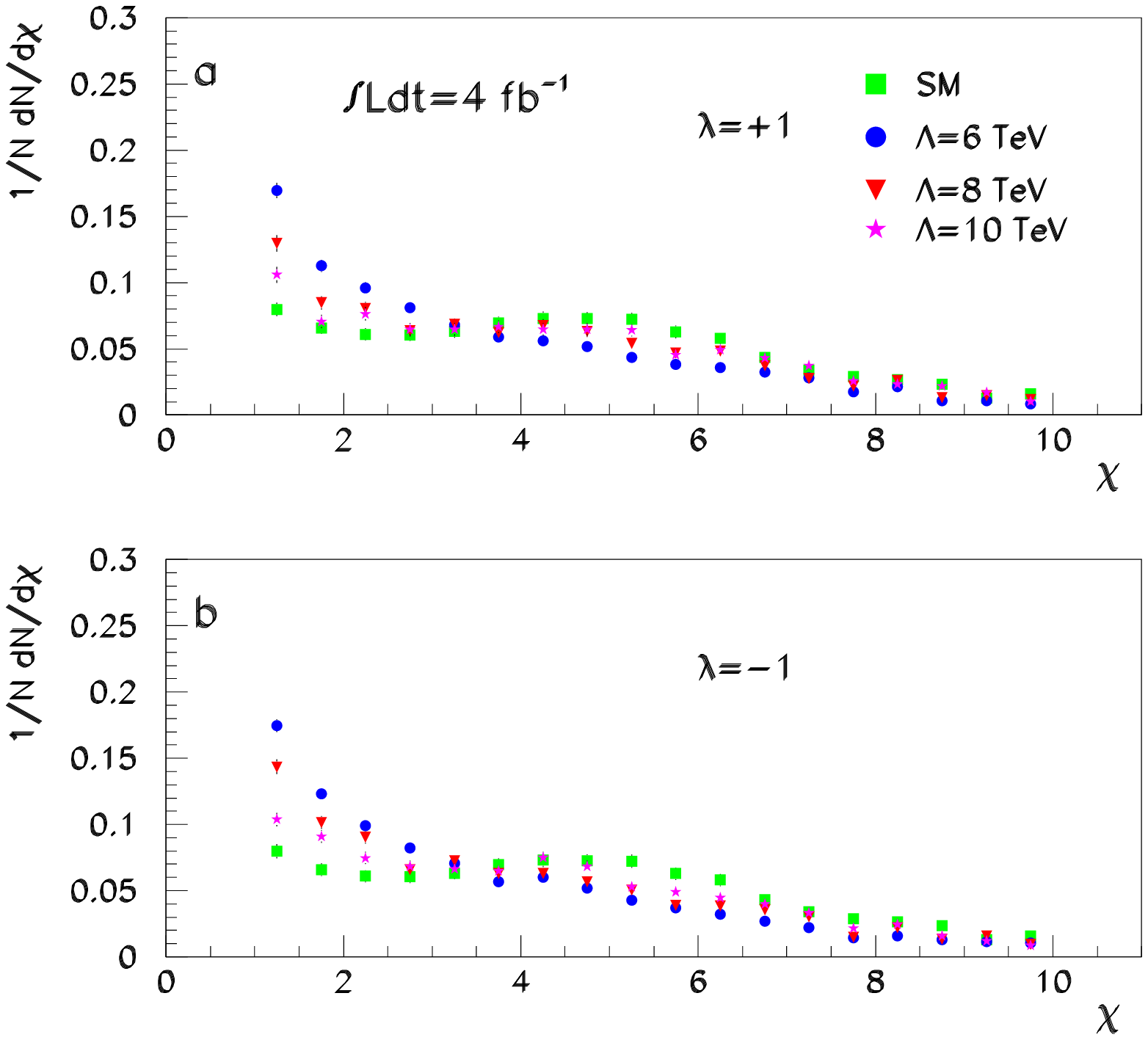}}
\vskip -2.0 cm
\caption{Dijet angular distributions for the Standard 
Model prediction compared to the quark    
contact interaction predictions at different 
energy scales $\Lambda$ and $E_{CM}=10\,$TeV. The integrated    
luminosity is assumed to be 4 fb$^{-1}$: a) destructive interference term; 
b) constructive interference term.}
\end{figure}

In fact, the variable ${\alpha}_{R}$  was introduced  
for dijet SUSY searches and QCD background suppression\cite{rand}.
We find that this kinematic variable  
can be used in the quark compositeness study as well.
The QCD induced $2 \to 2$ events have  ${\alpha}_{R}$ distribution
which sharply decreases  around $\sim 0.5.$                      
The ${\alpha}_{R}$
distribution of the events with the contact interaction
contribution has  different behaviour near $\sim 0.5$.

For our previous estimations of the compositeness scale $\Lambda$
at the LHC\cite{tdr} we considered the variable $R_{\chi}$\cite{cdf96}
\begin{equation}
   R_{\chi} = {N(\chi < {\chi}_0) \over N(\chi > {\chi}_0)},    
\end{equation}
which is the ratio of the number of dijet events with $\chi < {\chi}_0$ to
the number of dijet events with $\chi > {\chi}_0$.
This single parameter is helpful for describing the whole
$\chi$ distribution. The appropriate choice of ${\chi}_{0}$
maximizes the sensitivity of the variable $R_{\chi}$ to the 
features of the $\chi$ distribution.

The estimation of $\Lambda$ in this study is based on the 
analysis of ${\alpha}_{Z}$. As in   the case of the 
variable $\chi$, 
to describe  the whole distribution with a single parameter, 
we constructed  the variable
\begin{equation}
   R_{{\alpha}_{Z}} ={N({\alpha}_{Z}<{\alpha}_{Z}^{0}) \over
                N({\alpha}_{Z}>{\alpha}_{Z}^{0})},       
\end{equation}
where $N({\alpha}_{Z} > {\alpha}_{Z}^{0})$ 
($N({\alpha}_{Z} < {\alpha}_{Z}^{0})$) 
is the number of 
dijet events with ${\alpha}_{Z} > {\alpha}_{Z}^0$     
(${\alpha}_{Z} < {\alpha}_{Z}^0$).        

In order to know to what extent the observations   
either  conform or disprove the  quark compositeness 
scenario we consider the significance
\begin{equation}
   S = {{ R_{{\alpha}_Z}({\Lambda}) - R_{{\alpha}_Z}(SM)} 
              \over {\sigma}},    
\end{equation}
where $\sigma = \sqrt{ {{\sigma}^2_{\Lambda}}   + {{\sigma}^2_{SM}}}$.
               
It must be stressed that
the distribution of $|{\phi}_1-{\phi}_2|$, where
${\phi}_{1,2}$ are the azimuthal angles of the leading jets,             
for events with composite quarks show 
deviations from the SM prediction. This deviation depends
on the compositeness scale $\Lambda$ but is significantly
smaller  than that of the $\chi$,  ${\alpha}_R$ or $\alpha_Z$ 
distributions.

\section{Data simulation for generic LHC detector}
$ $

The simulation of the $pp$ collision with a quark    
substructure scenario was performed
with the event generator PYTHIA6.4\cite{pyth}. 
We employ the leading-order CTEQ6L1\cite{pump}  pdfs everywhere 
unless  otherwise stated
\footnote {We used PYTHIA6.4            
default choices for $Q^2$ definition as well as
factorization and renormalization scales}.
The initial-state and final-state 
QCD and QED radiation and multiple interactions were enabled.
We choose the case where  all quarks
are composite. The model with the left-left isoscalar contact 
interaction term is supposed\cite{hinc}.
The events were generated with the hard subprocess transverse      
momentum  $p_{T} > 1\,$TeV. 

\begin{figure}[ht]
\vskip -0.3cm
\centerline{\epsfxsize 5.0 truein \epsfbox{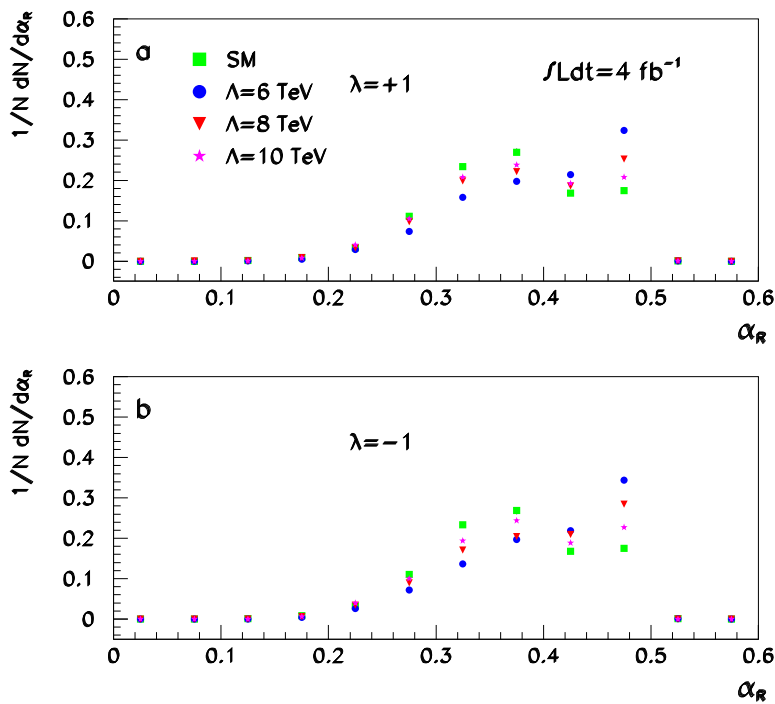}}
\vskip -2.0 cm
\caption{Normalized ${\alpha}_{R}$  distributions (see the text) 
for the Standard Model         
predictions compared to the quark contact interaction predictions 
at different energy scales $\Lambda$ and $E_{CM}=10\,$TeV.
The integrated luminosity is assumed to be 4 fb$^{-1}$: 
a) destructive interference term;$\,\,$
b) constructive interference term.}
\end{figure}

The detector performance was simulated by using the 
publicly available PGS-4\cite{conw} 
package written by J.~Conway and modified
by S.~ Mrenna for  the generic  LHC detector.                         
The calorimeter granularity is set to 
$(\Delta \phi \times \Delta \eta)=(0.10 \times 0.10)$. Energy smearing
in the hadronic calorimeter of the generic LHC detector is governed by\footnote
{We add  the constant term in the PGS-4 simulation of energy smearing
in the hadronic calorimeter}
\begin{equation}
  {\Delta E \over {E}} = {{0.6 \over \sqrt{E}} + 0.03} \qquad (E\,\,in\,\,GeV).
\end{equation}
Jets were reconstructed down to $|\eta|\le 3$ using 
the  $k_T$ algorithm implemented in PGS-4. We chose $ D=0.7$
for the jet resolution parameter and required that $E_{T}(jet)>100\,$GeV.
We use the simplified     
output from PGS-4, namely, a list of two most energetic
jets.
The average $P_T$
and invariant mass of the leading jets are
$<P_T^{j1}>=1.12\,$TeV, $<P_T^{j2}>=1.01\,$TeV, 
$<m_{j1j2}>=2.46\,$TeV, respectively.

\begin{figure}[ht]
\vskip -0.3cm
\centerline{\epsfxsize 5.0 truein \epsfbox{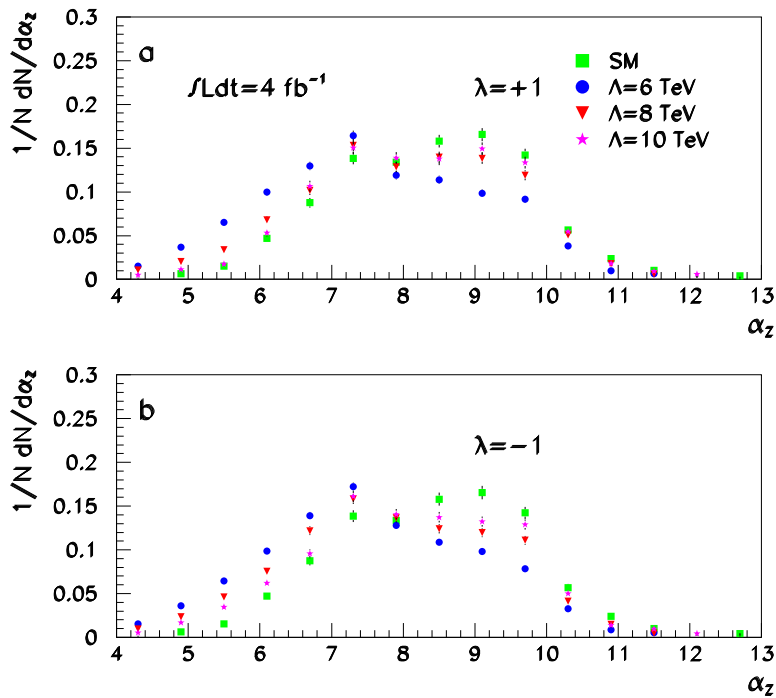}}
\vskip -2.0 cm
\caption{Normalized ${\alpha}_{Z}$  distributions (see the text) 
for the Standard Model         
predictions compared to the quark contact interaction predictions 
at different energy scales $\Lambda$ and $E_{CM}=10\,$TeV.
The integrated luminosity is assumed to be 4 fb$^{-1}$: 
a) destructive interference term;$\,\,$
b) constructive interference term.}
\end{figure}

\section{What we are going to look for}          
$ $

Measurements of particle production at a very high transverse
momentum at the LHC will provide a broad domain for new
physics discovery.   
It has already been noted that
the quark compositeness can be tested by
looking for deviations of the jet transverse momentum 
distribution from the QCD prediction.
In Fig.~1 we plot the $P_T$ distributions for two leading
jets predicted by the SM as well as by the quark substructure       
effect with  different compositeness scale $\Lambda$.
The data were simulated with
constructive $(\lambda=-1)$ interference term. Evidently, effect of            
quark substructure dominates over SM behaviour for high $P_T$
jets.

The  effect of quark compositeness is most     
pronounced in the high dijet mass region\cite{tdr}. For further
analysis  the events are chosen to have $m_{j1j2}>2.7\,$TeV.
After applying this cut  to  leading jets we have
$<P_T^{j1}>=1.27\,$TeV, $<P_T^{j2}>=1.18\,$TeV, 
$<m_{j1j2}>=3.23\,$TeV, respectively.
   
The comparison of the dijet angular distribution
predicted by the SM and induced by the quark substructure
with different values of $\Lambda$
is shown in Fig.~2.                                       
Plots $a$ and $b$ correspond 
to the  normalized $\chi$ distributions, $(1/N)(d N/d\chi)$,
for destructive and constructive interference,
respectively. 
More isotropic angular distribution in the  composite
quark interactions lead to the $\chi$ distribution
which peaked at low $\chi$.
\begin{figure}[ht]
\vskip -3.3cm
\centerline{\epsfxsize 6.5 truein \epsfbox{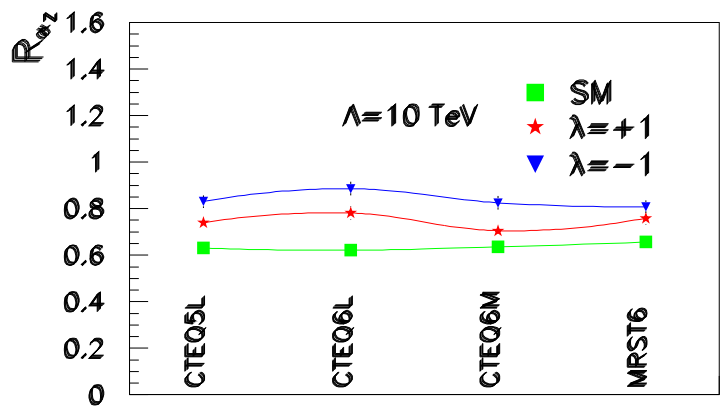}}
\vskip -6.0 cm
\caption{The dependence of $R_{{\alpha}_{Z}}$ (see the text) on the  
parton distribution functions.
The compositeness energy scale is set to be $\Lambda=10\,$TeV. 
The integrated luminosity
is assumed to be 4 fb$^{-1}$ and $E_{CM}=10\,$TeV.}
\end{figure}

The normalized ${\alpha}_{R}$ distributions  for destructive    
and constructive interference are demonstrated in Fig.~3 $a$ and $b$,
respectively.
As becomes apparent from Fig.~3,
quark compositeness leads to an enhancement
(abatement) in the distribution at $\alpha_R > 0.4\,(\alpha_R < 0.4)$
in comparison to the SM prediction.
In Fig.~4 we show the normalized ${\alpha}_{Z}$ distributions
for destructive (a) and constructive (b) interference.
The change in the behaviour  of the $\alpha_Z$ distributions 
in the vicinity  of 
$\alpha_Z = 8$ is   quite impressive.
The data sensitivity in Figs.~2, 3 and 4 corresponds to the 
integrated luminosity of 4 fb$^{-1}$.

The distributions of the variables $\chi$, $\alpha_R$ and
$\alpha_Z$ show a robust signal and can be used independently
to observe effects from the quark substructure.
One can see that the effect from quark compositeness is higher     
for constructive interference than for destructive one.

One of the main uncertainties in the interpretation of 
new physics at the LHC detectors will come from pdfs. 
Unfortunately, the pdfs are not so well determined in the 
kinematic   region with high transverse momentum  jets.
Figure~5  shows the effect of the pdfs choice on the differences
between the SM and composite quarks predictions for $R_{{\alpha}_Z}$.
The result obtained with the leading-order CTEQ6L1\cite{pump}    
parametrization 
is compared with the results which correspond to the
leading-order CTEQ5L1\cite{pump} and 
next-to-leading order CTEQ6M1\cite{pump} and MRST 2006\cite{mart} pdfs.
We can conclude that the LHC potential for the quark substructure
study can be slightly affected by pdfs uncertainties.

In Fig.~6 we show the LHC lower limits on 
$\Lambda$ up to which the effects of 
quark compositeness can be observed at E$_{CM}=10\,$TeV
as a function of the 
 accumulated luminosity.
The data correspond  to the significance of eq.(7) close
to S=3, for which we can claim that we have strong evidence 
for the observed signal.    
The parameter ${\alpha}_{Z}^{0}$
used here is ${\alpha}_{Z}^{0}=8.0.$
The data were obtained with
the inclusion of statistical uncertainties alone. 
We  performed analogous analysis   of lower limits
on $\Lambda$ with the use of the variables $\chi$ and $\alpha_R$.
The estimates    of $\Lambda$ agree with those obtained
with the use   of $\alpha_Z$ within
the $\pm 0.5\,$TeV gate.
We find that 
with the LHC running at the design center-of-mass energy    
$E_{CM}=14\,$TeV,  one will have
a gain  of $\sim 0.5 \,$TeV to the values of $\Lambda$           
obtained for $E_{CM}=10\,$TeV at the integrated luminosity
in the range from 0.5 to 10 fb$^{-1}$.
For this  analysis we used  $m_{j1j2}>3.0\,$TeV.

\section{Conclusions }
$ $

We examined the capability of the LHC experiments to observe
quark substructure in the early stages of the running 
considering that the LHC would have an accumulated
luminosity in the range from  0.5 to 10.0 fb$^{-1}$ 
at $E_{CM}=10\,$TeV.

\begin{figure}[ht]
\vskip -3.3cm
\centerline{\epsfxsize 6.5 truein \epsfbox{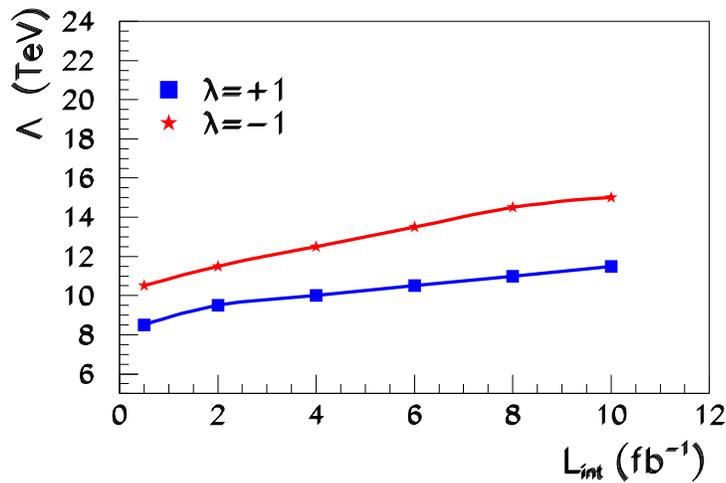}}
\vskip -5.5 cm
\caption{LHC reach on  the characteristic energy scale 
$\Lambda$  as a function of the integrated luminosity 
for constructive ($\lambda = -1$) and destructive ($\lambda = +1$) interference term  
and $E_{CM}=10\,$TeV.}
\end{figure}
We show that quark compositeness at the LHC              
can be manifested by different dijet kinematic variables.
The distribution of two novel variables ${\alpha}_R$\cite{rand}
and $\alpha_{Z}$
might provide direct hint to the quark substructure
observation as well as dijet transverse momenttum 
and angular distributions.
Analysis of the variable ${\alpha}_Z$ defined in the paper
can be effective for  estimation 
of the compositeness energy scale $\Lambda$.
In the early stages of the LHC running
it is possible to probe the compositeness scale $\Lambda$ 
up to 10.5(15.0) TeV for the constructive interference term
and 8.5(11.5) TeV for the destructive one 
for the  integrated luminosity
of 0.5(10) fb$^{-1}$ and $E_{CM}=10\,$TeV.

The last but not the least,  such a features of the transverse momenta
and angular distribution for high energy jets 
can have different new physics sources 
(SUSY, extra dimensions, Technicolor etc.). In this case 
the use of combinations of observables will be most effective
for the discrimination and 
interpretation of the data and the understanding of the underlying
theory.

{ \small {
      
}}
} 

\end{document}